\documentclass[aps,prb,twocolumn,superscriptaddress,showpacs,floatfix]{revtex4-1}
\usepackage{amsmath,amssymb,graphicx}
\usepackage{wasysym}
\usepackage[utf8]{inputenc}
\usepackage[T1]{fontenc}
\usepackage{xcolor}
\IfFileExists{newtxtext.sty}
   {\usepackage{newtxtext,newtxmath}}
   {\IfFileExists{stix.sty}
      {\usepackage{stix}}
      {\IfFileExists{mathptmx.sty}
      {\usepackage{mathptmx}}{} } }
\usepackage{textcomp}

\usepackage{bm}

\IfFileExists{siunitx.sty}{\usepackage{booktabs,siunitx}}{}

\pdfoutput=1
\usepackage{color}
\definecolor{LinkColor}{rgb}{0.256,0.439,0.588}

\usepackage{pifont}

\usepackage{hyperref}
\hypersetup{
	colorlinks = true,
	linkcolor = [rgb]{0.70,0.13,0.13},
	citecolor = [rgb]{0.13,0.55,0.13},
	urlcolor = [rgb]{0.25,0.41,0.88}}

\DeclareSymbolFont{sfletters}{OML}{cmbrm}{m}{it}
\DeclareMathSymbol{\sfeps}{\mathord}{sfletters}{"22}

\graphicspath{{Figures/}}

\begin{document}
\title{Vestigial anyon condensation in kagome quantum spin liquids}

\author{Yan-Cheng Wang}
\affiliation{School of Materials Science and Physics, China University of Mining and Technology, Xuzhou 221116, China}

\author{Zheng Yan}
\affiliation{Department of Physics and HKU-UCAS Joint Institute of Theoretical and Computational Physics, The University of Hong Kong, Pokfulam Road, Hong Kong SAR, China}
\affiliation{State Key Laboratory of Surface Physics, Fudan University, Shanghai 200433, China}

\author{Chenjie Wang}
\affiliation{Department of Physics and HKU-UCAS Joint Institute of Theoretical and Computational Physics, The University of Hong Kong, Pokfulam Road, Hong Kong SAR, China}

\author{Yang Qi}
\email{qiyang@fudan.edu.cn}
\affiliation{State Key Laboratory of Surface Physics, Fudan University, Shanghai 200433, China}
\affiliation{Center for Field Theory
	and Particle Physics, Department of Physics, Fudan University,
	Shanghai 200433, China}
\affiliation{Collaborative Innovation Center of Advanced
	Microstructures, Nanjing 210093, China}

\author{Zi Yang Meng}
\email{zymeng@hku.hk}
\affiliation{Department of Physics and HKU-UCAS Joint Institute of Theoretical and Computational Physics, The University of Hong Kong, Pokfulam Road, Hong Kong SAR, China}

\begin{abstract}
We construct a lattice model of topological order (Kagome quantum spin liquids) and solve it with unbiased quantum Monte Carlo simulations. A three-stage anyon condensation with two transitions from a $\mathbb Z_2\boxtimes\mathbb Z_2$ topological order to a $\mathbb Z_2$ topological order and eventually to a trivial symmetric phase is revealed. These results provide concrete examples of phase transitions between topological orders in quantum magnets. The designed quantum spin liquid model and its numerical solution offer a playground for further investigations on vestigial anyon condensation.
\end{abstract}

\date{\today}
\maketitle

\section{Introduction} Quantum spin liquids (QSLs)~\cite{YiZhou2017,Broholm2020} are the embodiment of topological orders and offer the ideal platform for systematical investigations of  fractional anyonic excitations and statistics therein~\cite{Wen2017,Wen2019}. While  experimental progress on QSL and topological orders is difficult and often hampered by the complexity of materials and limitation of probing techniques, such as how to remove the impurity scattering of kagome antiferromagnets herbertsmithite and Zn-doped barlowite~\cite{HanTH12,FengZL17,WeiYuan2017,ZLFeng2019,JJWen2019,YuanWei2020,YuanWei2020nano}, theoretical progress on both topological orders and quantum phase transitions between them is fast and promising. However, most theoretical studies on topological phase transitions either stay at the algebraic level of anyon condensation~\cite{Bais2002,Bais2009,Burnell2018}, or are based on perturbed exactly solvable yet unrealistic models such as string-net models~\cite{Levin2005,Gilis2009,Burnell2011,Burnell2016,Schulz2013,Schotte2019,Marien2017,Wiedmann2020}.   Realistic models and their fully quantum many-body solutions of topological orders are rare, for obvious reasons: the lack of imagination in model construction and the lack of unbiased numerical methodology to handle these correlated systems.

Here, we hit two birds with one stone. By designing a lattice model of coupled kagome QSLs that involve only two-spin interactions and solving it with large scale quantum Monte Carlo (QMC) simulations, amended with an analysis on anyon condensation transitions, we found that our model offers a three-stage anyon condensation process, of a vestigial type~\cite{Nie2014}, from a  $\mathbb Z_2\boxtimes\mathbb Z_2$ topological order QSL to a $\mathbb Z_2$ topological order QSL and eventually to a trivial symmetric phase. The difference in the two topological orders, dubbed QSL-I and QSL-II, lies in their underlying anyon excitations, which we reveal with topological Wilson loops that signal anyon condensation and dynamical spin spectra that exhibit spinon confinement from QSL-I to QSL-II. The phase transitions between the two QSLs and into other phases are scrutinized, with the nature of the transition, first order versus continuous, and the type of anyon condensation, symmetry breaking, and universality class clarified from topological field theory analysis and unbiased quantum many-body numerics.

\section{Model and Method} 
\subsection{Model} 
\begin{figure*}[htp!]
	\centering
	\includegraphics[width=\textwidth]{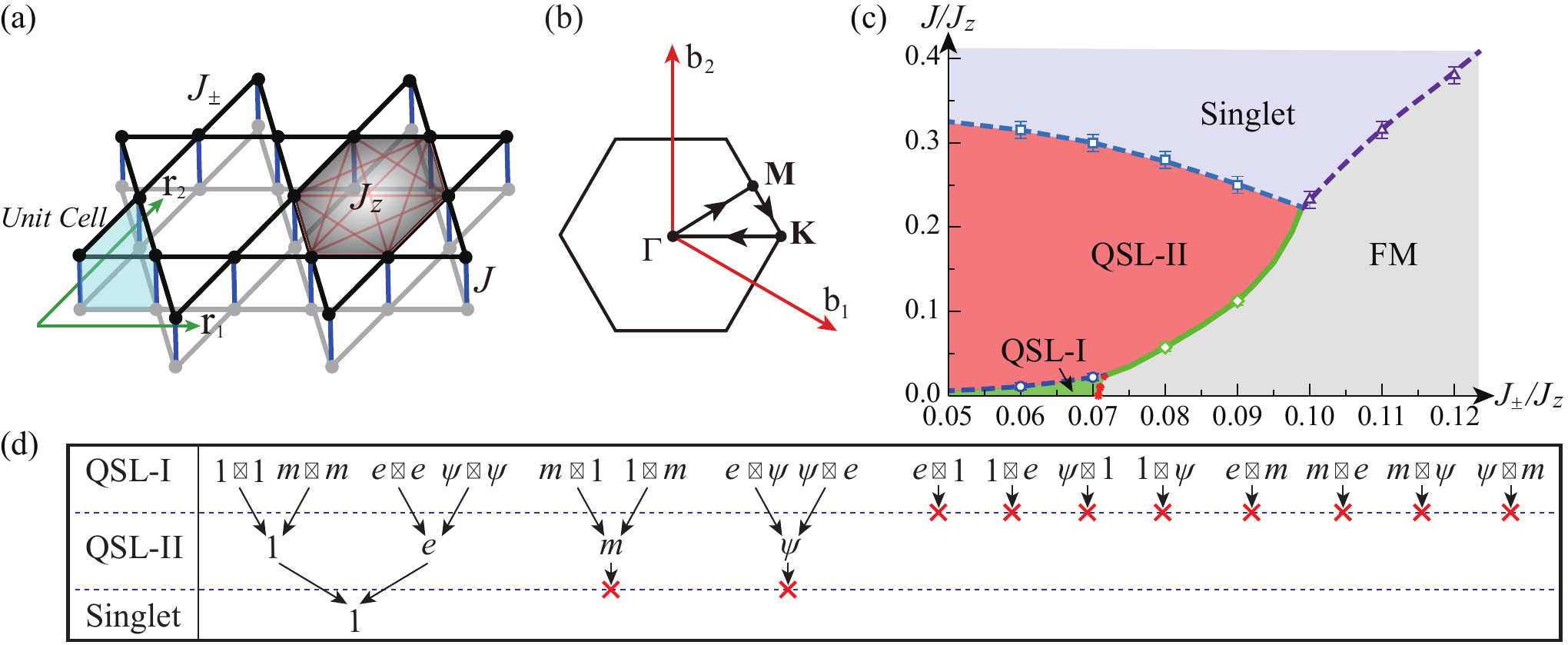}
	\caption{(a) Bilayer kagome lattice model with the six-site unit cell and lattice vectors $\mathbf{r}_{1,2}$. The nearest-neighbor ferromagnetic (FM) in-plane transversal interaction $J_{\pm}$ (black bonds), the hexagonal antiferromagnetic in-plane longitudinal interaction $J_{z}$ (gray-shaded hexagon), and the interlaryer antiferromagnetic Heisenberg interaction $J$ (blue bonds) are present. (b) Brillouin zone of the bilayer kagome lattice, with the reciprocal space vectors $\mathbf{b}_1$ and $\mathbf{b}_2$ and the high-symmetry points $\Gamma$, $M$, and $K$.  (c) QMC phase diagram of the model spanned by the axes of $J_{\pm}/J_{z}$ and $J/J_{z}$. The two quantum spin liquids (QSL-I and QSL-II) and the FM and singlet phases are shown. Dashed lines are continuous phase transitions and the solid line is the first-order phase transition. Symbols represent the places where the QMC parameter scans are performed. (d) Schematic process of the vestigial anyon condensation from QSL-I to QSL-II and eventually to the trivial singlet phase. 1, $e$, $m$, and $\psi$ are the four anyons of $\mathbb Z_2$ topological order and red X's represent the confinement of the corresponding anyons at the phase transition.}
	\label{fig:fig1}
\end{figure*}

We design a lattice model that hosts a  $\mathbb Z_2\boxtimes\mathbb Z_2$ kagome QSL phase and solve it with large-scale QMC simulations. As shown in Fig.~\ref{fig:fig1} (a), the model lives on a bilayer kagome lattice with a six-site unit cell and Hamiltonian reads
\begin{equation}
H =-J_{\pm}\sum_{\langle i,j \rangle} (S_i^{+}S_j^{-}+\text{H.c.}) + \frac{J_z}{2}\sum_{\hexagon}\Big(\sum_{i\in\hexagon} S_i^z\Big)^{2} +J\sum_{\langle i,j \rangle^{'}} \mathbf{S}_i\cdot\mathbf{S}_j 
\label{eq:eq1}
\end{equation}
where $J_{\pm}$ is the ferromagnetic (FM) transverse nearest-neighbor interaction in the kagome plane, $J_{z}$ is the antiferromagnetic longitudinal interactions between any two spins in the hexagon of the kagome plane, and $J$ is the interlayer antiferromagnetic Heisenberg interaction.
Throughout the paper, we set $J_{z}=1$ as the energy unit.
The first two terms constitute the two-layered Balents-Fisher-Girvin (BFG) model~\cite{BFG2002}, and are coupled together by the last term.
Since the BFG model in each layer only has $U(1)$ symmetry, even if the interlayer interaction is SU(2) symmetric, the global spin rotational symmetry of the system is still $U(1)$. We stress that our bilayer BFG model is the first one that realizes anyon condensation and that involves only two-spin interaction.

\subsection{Quantum Monte Carlo method}
To investigate the ground state phase diagram of Eq.~(\ref{eq:eq1}), we employ large-scale stochastic series expansion (SSE) QMC simulations~\cite{Syljuaasen2002,Sandvik2010} with a plaquette update and generalized balance condition~\cite{YCWang2017,YCWang2018,YCWang2018SET}. 
As for the spectral functions, we employ the stochastic analytic continuation method~\cite{Sandvik1998a,Beach2004,Syljuasen2008,Sandvik2015,Qin2017,Shao2017,GYSun2018,NvsenMa2018a,YanZheng2020,ChengkangZhou2020}  to obtain the real frequency spin excitation spectra from the QMC imaginary-time correlation functions. 

\begin{figure}[htp!]
	\centering
	\includegraphics[width=\columnwidth]{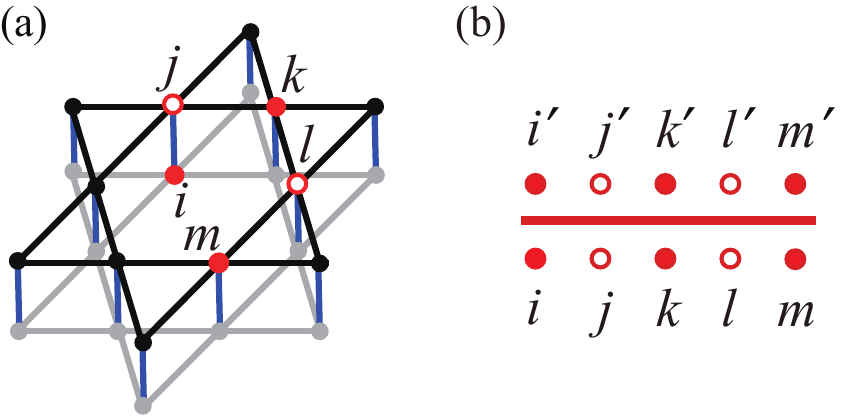}
	\caption{Plaquette decomposition with five sites and ten legs in a vortex.}
	\label{fig:figs1}
\end{figure}
To overcome the strong frustration of the system, we take the plaquette decomposition with five sites and ten legs in a vortex of the Hamiltonian, Eq.~(\ref{eq:eq1}) , which are shown as Fig.~\ref{fig:figs1}.
Here, in each bilayer hexagon, we consider both $C_6$ and layer-inversion symmetries of the lattice and make sure the five-site plaquette unit fully covers all sites of the lattice and all the interactions of the Hamiltonian. Then Eq.~(\ref{eq:eq1}) is decomposed into diagonal operators,
\begin{eqnarray}
H_{\text{diag}} = C & - & \frac{J_z}{z_1} \left(S_j^{z}S_k^{z}+S_k^{z}S_l^{z}+S_l^{z}S_m^{z}\right)\nonumber\\
& - &  \frac{J_z}{z_2}\left(S_j^{z}S_l^{z}+S_k^{z}S_m^{z}+S_j^{z}S_m^{z}\right)-\frac{J}{z_3}(S_i^{z}S_j^{z}),
\label{eq:h1b}
\end{eqnarray}
where $z_1=3$, $z_2=2$, and $z_3=4$ are prefactors to avoid over-counting of bonds, and off-diagonal operators:
\begin{eqnarray}
H_{\text{off-diag}} &=&\frac{J_{\pm}}{z_1} \left(S_j^{+}S_k^{-}+S_k^{+}S_l^{-}+S_l^{+}S_m^{-} + h.c. \right)\nonumber\\
         &+&\frac{J}{z_3}\left(S_i^{+}S_j^{-} + h.c. \right).
\label{eq:h2b}
\end{eqnarray}

\begin{figure}[htp!]
	\centering
	\includegraphics[width=\columnwidth]{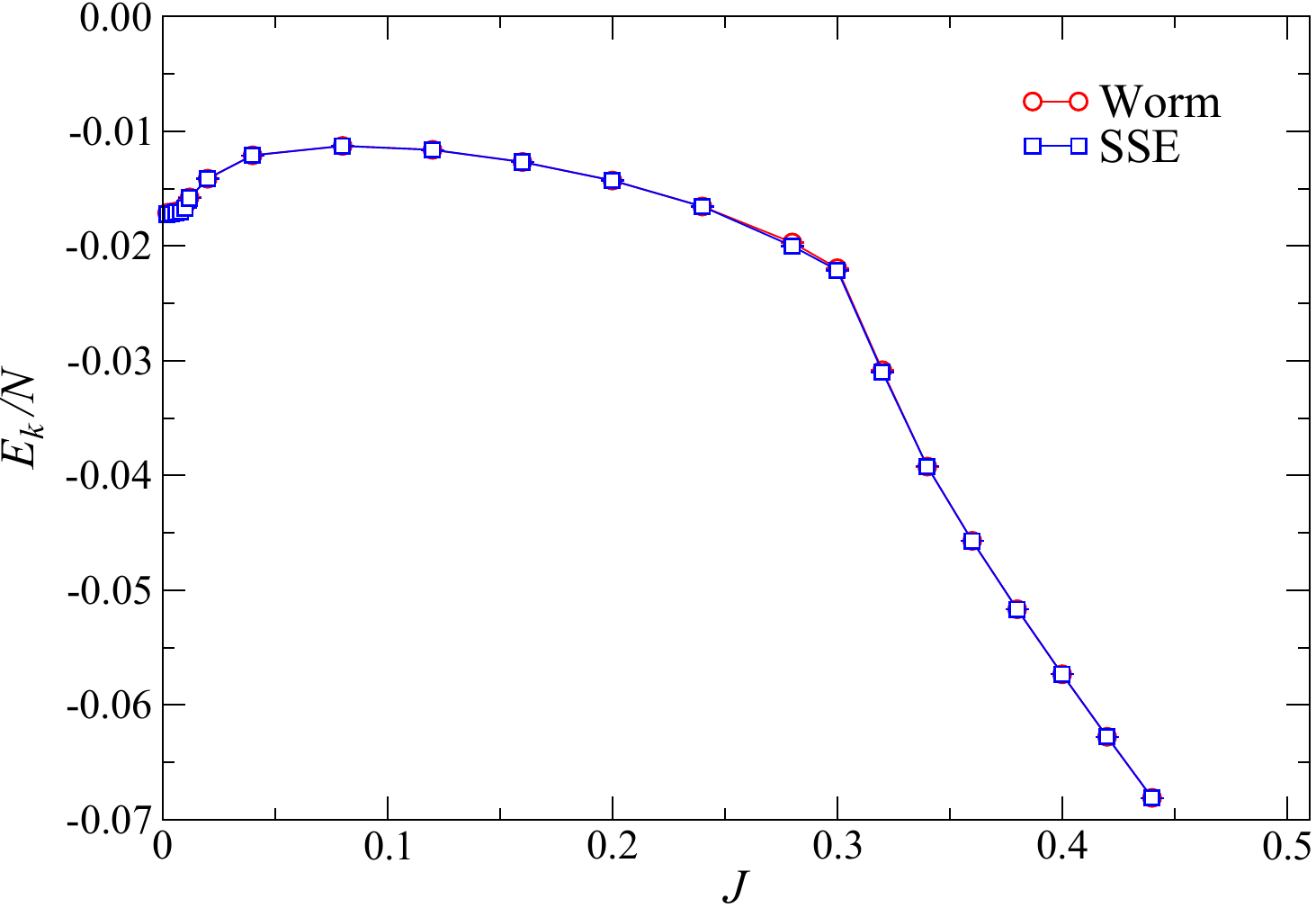}
	\caption{The kinetic energy density $E_k/N$ as a function of $J$ at $J_{\pm}=0.06$ with system size $L=12$ and the inverse of temperature $\beta J_{\pm} =2L$ calculated with both the SSE-QMC and the worm QMC; the results are identical.}
	\label{fig:figs2}
\end{figure}

After the decomposition, we implement the Monte Carlo method with the general balance condition without detail balance~\cite{SuwaPRL2010} to obtain the solution of probability equations which strongly reduces the unpreferred bounce update~\cite{YCWang2017QSL,YCWang2018}.

To benchmark the SSE-QMC code, we also implemented a worm-type continuous-time QMC ~\cite{Prokofev1998JETP,Prokofev1998PLA} for the same model.
As shown in Fig.~\ref{fig:figs2}, we used these two methods to calculate the kinetic energy density $E_k/N$ as a function of $J$ at $J_{\pm}=0.06$ with system size $L=12$ and inverse temperature $\beta J_{\pm}=2L$, and obtained identical results. The more important reason for implementing the worm-QMC is that in the worm representation, the transverse dynamical spin correlation function $S^{\pm}_{\alpha\beta}(\mathbf{q},\tau)$ is easier to measure as the head (or tail) of the worm is born with the "exact" imaginary time, whereas the longitudinal dynamical spin correlation $S^{zz}_{\alpha\beta}(\mathbf{q},\tau)$ is easier to measure with the SSE-QMC since it is a diagonal measurement. This is how the data on dynamical spin correlation functions in Fig.~\ref{fig:fig4}  were obtained.

\section{Measurements and Results}
\subsection{Kinetic energy} 
\begin{figure}[htp!]
	\centering
	\includegraphics[width=\columnwidth]{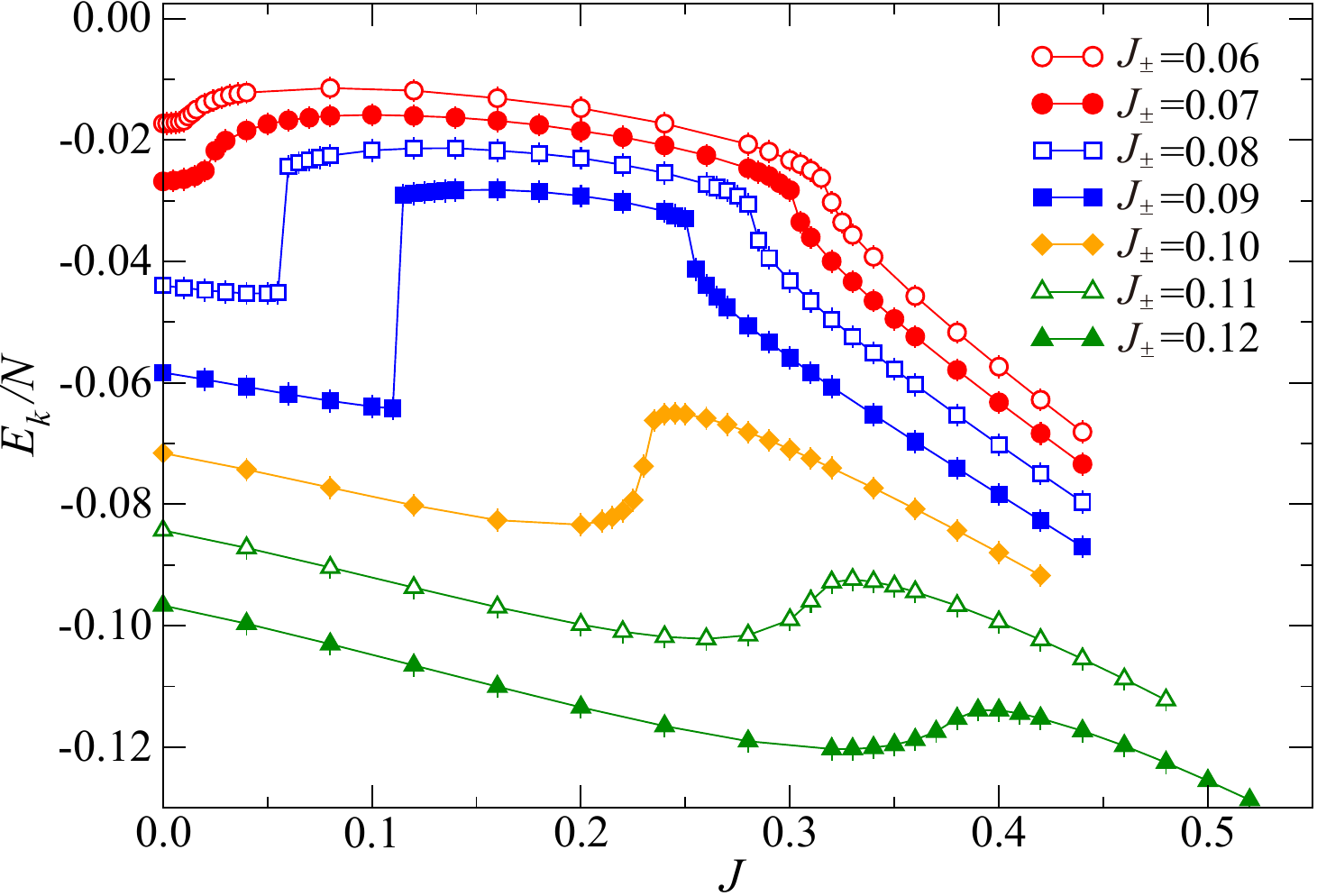}
	\caption{The density of the kinetic energy $E_{k}/N$ as a function of $J$ for $J_{\pm}=0.06, 0.07, 0.08, 0.09, 0.10, 0.11$, and $0.12$, respectively, with system size $L=18$ and inverse of temperature $\beta J_{\pm} =2L$.}
	\label{fig:fig2}
\end{figure}

To obtain the groundstate phase diagram as shown in Fig.~\ref{fig:fig1} (c), we first plot the kinetic energy density $E_{k}/N= \langle -\sum_{\langle i,j \rangle} (S_i^{+}S_j^{-}+\text{H.c.}) \rangle/N$, which is the expectation value of the transverse part of the Hamiltonian, Eq.~\eqref{eq:eq1}, with lattice size $N=6\times L^2$, linear size $L=18$, and inverse temperature $\beta=2L/J_{\pm}$. The results are shown in Fig.~\ref{fig:fig2}. We fix different initial values of $J_{\pm}$ and scan the value of $J$ to monitor how the $E_k/N$ values behave. At $J_{\pm}=0.06$ and $0.07$, where the single-layer kagome model is in a $\mathbb Z_2$ QSL phase as shown in previous work~\cite{Isakov2006}, the kinetic energy density demonstrates a turning point around $J=0.01$ and $0.022$ in a continuous manner. As shown schematically in Fig.~\ref{fig:fig1} (d), this is the anyon condensation transition between the $\mathbb Z_2\boxtimes\mathbb Z_2$ topological order QSL and the $\mathbb Z_2$ topological order QSL. With a further increase to  $J_{\pm}=0.08$ and $0.09$,  a discontinuous jump appears in the $J$ scans and this is the first-order phase transition from the FM phase to  QSL-II in the phase diagram in Fig.~\ref{fig:fig1} (c). The data on spin stiffness across these transitions are shown in Fig.~\ref{fig:figs3}. It is interesting to observe that, for the $J_{\pm}=0.06$, 0.07, 0.08, and 0.09 curves, there exists another turning point at larger values of $J$, which signifies the second anyon condensation transition from QSL-II to a trivial product state of interlayer singlets. This condensation transition is again schematically shown in Fig.~\ref{fig:fig1} (d). At the larger values of $J_{\pm}=0.10, 0.11$, and $0.12$, there exists only one transition between the FM phase and the singlet phase.

\subsection{Spin stiffness} 
\begin{figure}[htp!]
	\centering
	\includegraphics[width=\columnwidth]{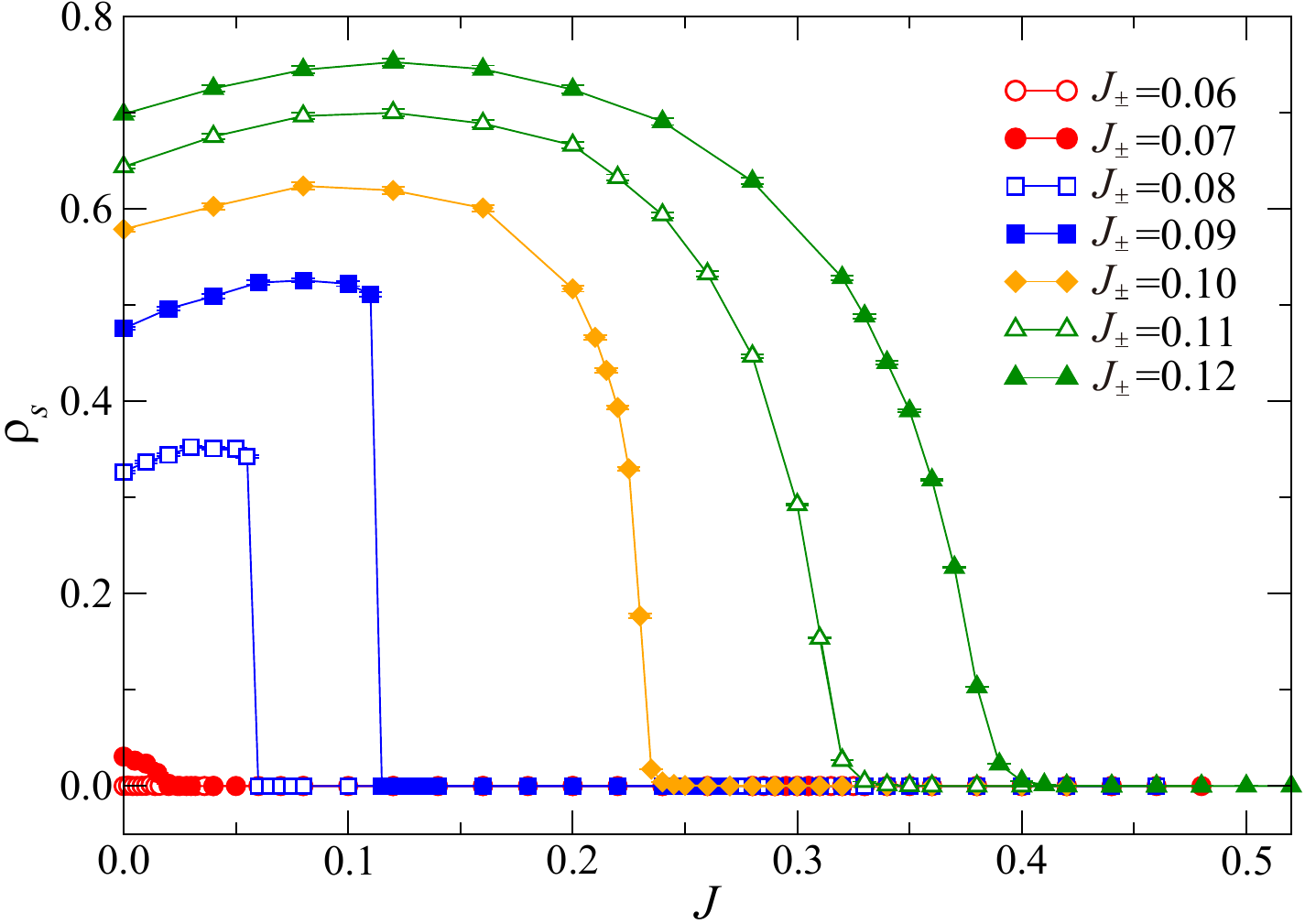}
	\caption{The spin stiffness $\rho_{s}$ as a function of $J$ for $J_{\pm}=0.06, 0.07, 0.08, 0.09, 0.10, 0.11$, and $0.12$ with system size $L=18$ and inverse of temperature $\beta J_{\pm} =2L$.}
	\label{fig:figs3}
\end{figure}

\begin{figure}[h!]
	\centering
	\includegraphics[width=\columnwidth]{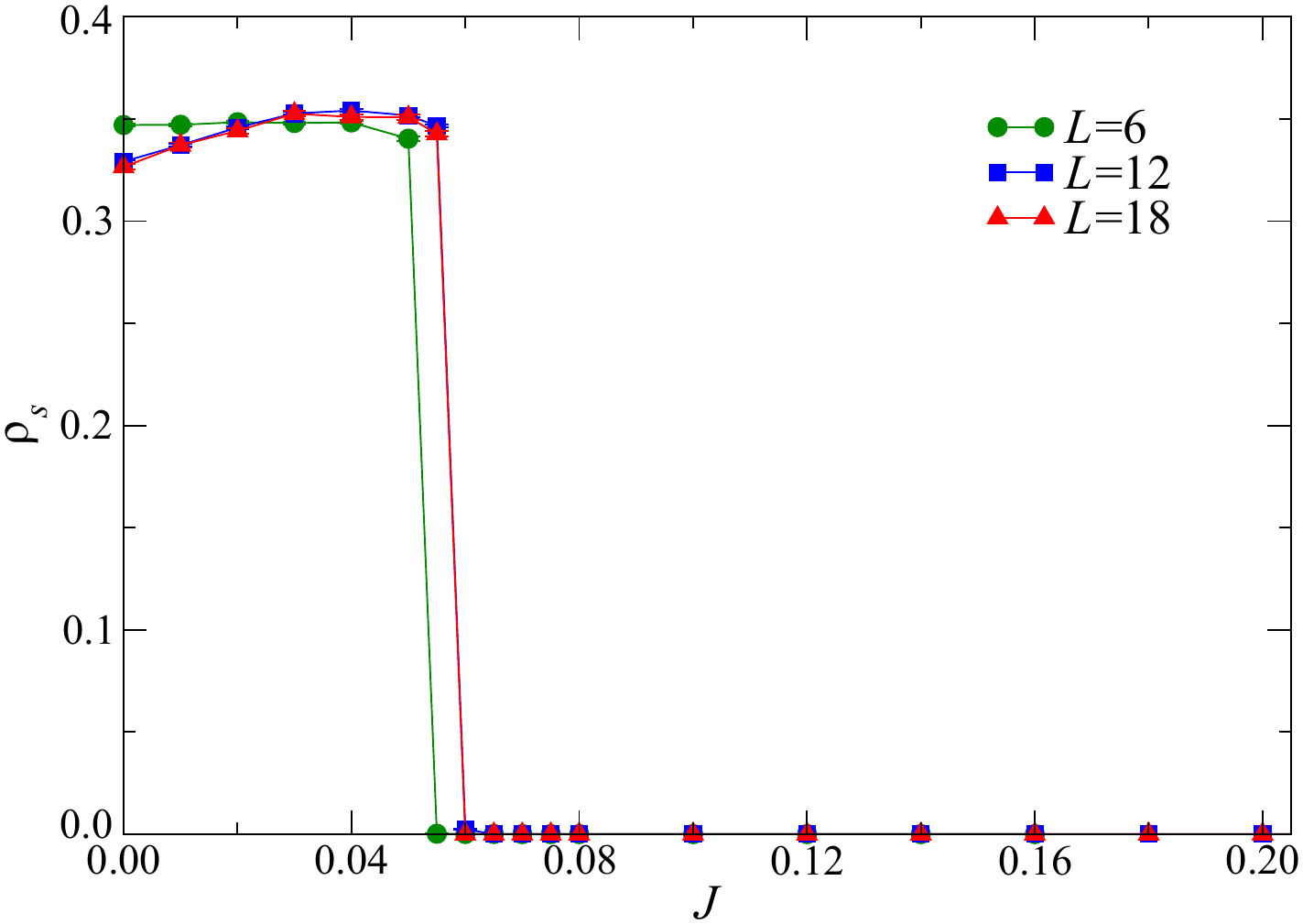}
	\caption{The spin stiffness $\rho_{s}$ as a function of $J$ for $J_{\pm}=0.08$ with system size $L=6, 12$, and $18$ and inverse of temperature $\beta J_{\pm} =2L$.}
	\label{fig:figs4}
\end{figure}

In order to construct the phase diagram of the model shown in Fig.~\ref{fig:fig1} (c), we also compute other observables such as the spin stiffness
 $\rho_s=(W_{\mathbf{r}_1}^2+W_{\mathbf{r}_2}^2)/(4\beta J_{\pm})$ through winding number fluctuations $W_{\mathbf{r}_{1,2}}^2$~\cite{Pollock1987}, where $\mathbf{r}_{1,2}$ are the two lattice directions.
The results of the spin stiffness $\rho_{s}$ as a function of $J$ for different $J_{\pm}$ values with system size $L=18$ are shown in Fig.~\ref{fig:figs3}. At $J_{\pm}=0.06$, $\rho_s=0$ for all values of $J$, meaning that the system is always in the QSLs and the singlet phase, without transverse long-range order. Only when $J_{\pm}=0.08$ and $0.09$ is the spin stiffness finite at small $J$ which means that the system is in the FM phase. However, when $J$ increases to the transition point, $\rho_s$ decreases sharply to 0, which reveals the first-order transition from the FM phase to QSL-II. We also consider the finite-size effect of this transition, as shown in Fig.~\ref{fig:figs4}, and find that when $L\ge 12$, it is then large enough to eliminate finite-size effects. Going back to Fig.~\ref{fig:figs3}, when $J_{\pm}\ge 0.10$, the $J$ scans again show a continuous drop, this is the three-dimensional O(3) transition from the FM phase to the singlet phase.

\begin{figure}[h!]
	\centering
	\includegraphics[width=\columnwidth]{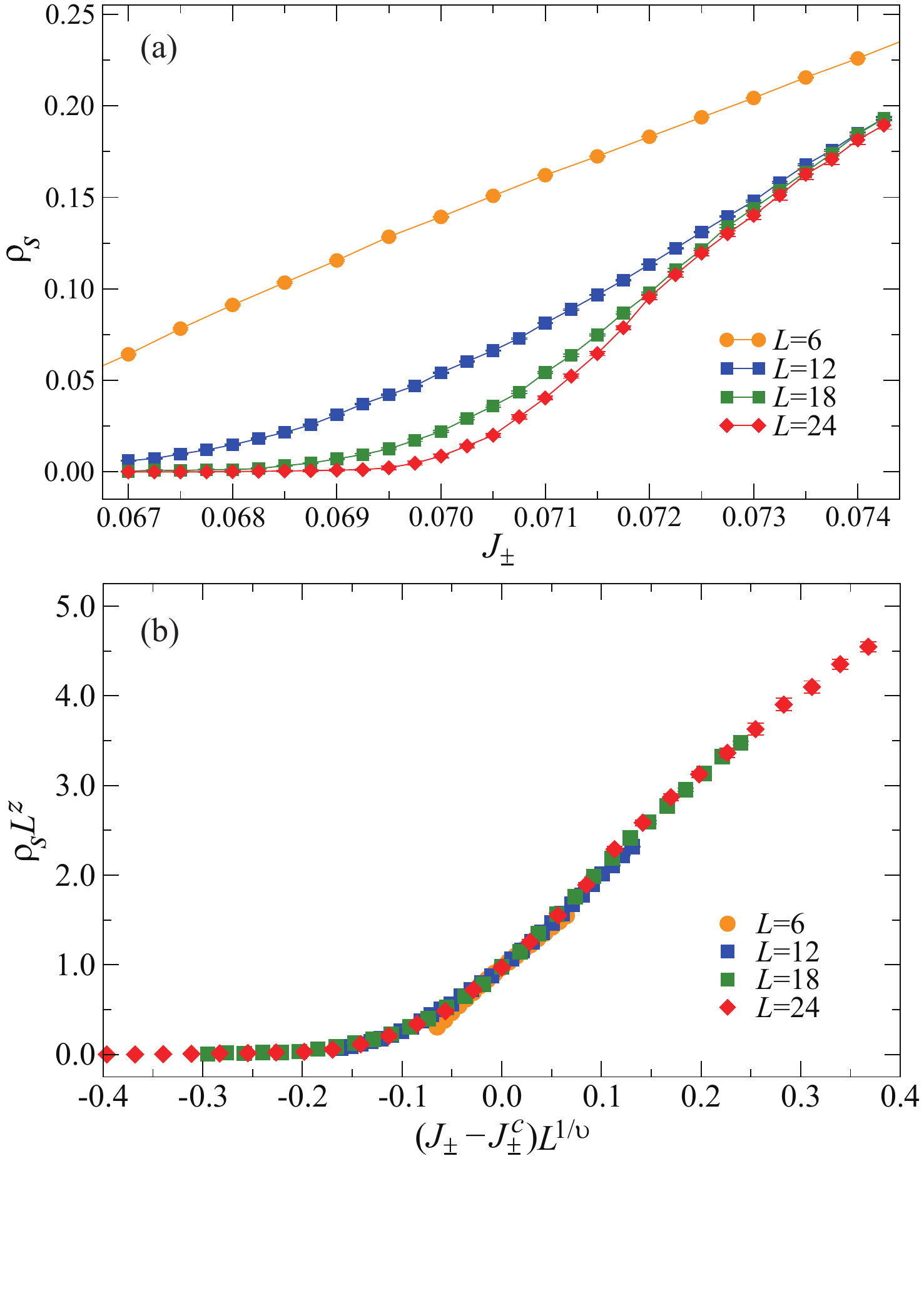}
	\caption{(a) The spin stiffness $\rho_{s}$ as a function of $J_{\pm}$ for $J=0.01$ with system size $L=6, 12, 18$, and $24$ and inverse of temperature $\beta J_{\pm} =2L$. (b) Data collapse of the spin stiffness  $\rho_{s}L^{z}$ as a function of $(J_{\pm}-J_{\pm}^{c}) L^{1/\nu}$.}
	\label{fig:figs5}
\end{figure}

To illustrate the phase transtion between QSL-I and the FM phase as shown in Fig.~\ref{fig:fig1} (c), we also simulate the spin stiffness as a function of $J_{\pm}$ for $J=0.01$ with system size $L=6, 12, 18$, and $24$, as shown in Fig.~\ref{fig:figs5}~(a), and then take the data collapse of the spin stiffness  $\rho_{s}L^{z}$ as a function of $(J_{\pm}-J_{\pm}^{c}) L^{1/\nu}$  with critical point $J_c=0.0710(4)$ and exponents $z=1$ and $\nu=0.67(1)$, as shown in Fig.~\ref{fig:figs5}~(b). This suggests that the transition from QSL-I to the FM phase is a continuous phase tansition  of 3D $XY^*$ universality.

\subsection{Wilson loop} Next, we focus on the two QSLs, and reveal the main discovery of this work: the vestigial anyon condensation.
We first recall that the single-layer BFG model~\cite{BFG2002} realizes $\mathbb Z_2$ topological order~\cite{Isakov2006,GYSun2018,Becker2018}, with four types of anyons:
the trivial anyon 1, the bosonic spinon $e$, the vison $m$, and the fermionic spinon $\psi$, which is obtained by fusing $e$ and $m$.
The operators $S^{\pm}$ and $S^z$ in Hamiltonian ~\eqref{eq:eq1} create/annihilate a pair of $e$ and $m$ anyons, respectively.
Hence, their spectra reflect the two-particle continuum of the corresponding anyons, and the two operators can be used to construct Wilson-loop operators, which we use below.
Furthermore, $e$ and $m$ anyons both carry nontrivial symmetry fractionalization: $e$ anyons carry half-integer spins and $m$ anyons carry a fractionalized crystalline momentum~\footnote{To be precise, they carry a projective representation of translation symmetries where $T_1T_2=-T_2T_1$.}.
Consequently, condensing either type of anyon leads to spontaneous breaking of the corresponding symmetries: condensing $e$ ($m$) anyons gives a continuous transition to an FM phase (a valence-bond-solid phase)~\cite{GYSun2018}, respectively.

\begin{figure}[htp!]
	\centering
	\includegraphics[width=\columnwidth]{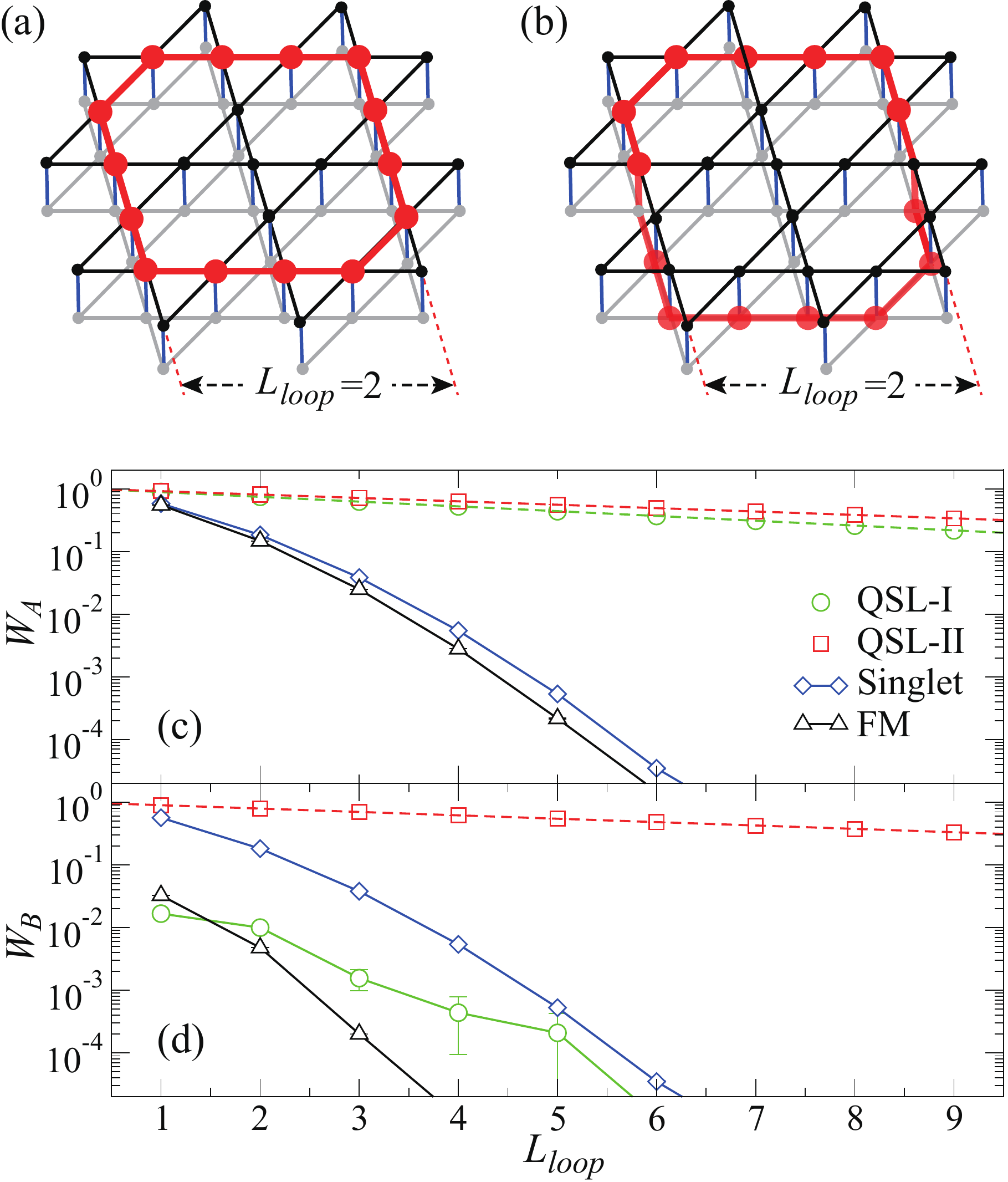}
	\caption{Two types of  Wilson loops $W_{A(B)}$  as a function of the side length of the loop $L_{loop}$ (proportional to the perimeter of the encircled region $M$) in different phases: QSL-I $(J_{\pm}=0.06$, $J=0.004)$, QSL-II $(J_{\pm}=0.06$, $J=0.20)$, Singlet $(J_{\pm}=0.06$, $J=0.40)$, and FM $(J_{\pm}=0.09$, $J=0.06)$, repectively, with system size $L=18$ and inverse of temperature  $\beta J_{\pm} =2L$.}
	\label{fig:fig3}
\end{figure}

In our bilayer model, the QSL-I phase is smoothly connected to the $J=0$ limit where the two layers decouple.
Hence, this topological order is a stacking of two $\mathbb Z_2$ topological orders.
We abuse the notation of the Deligne product $\boxtimes$ to represent such stacking and denote this topological order $\mathbb Z_2\boxtimes\mathbb Z_2$.
Anyon excitations in this topological order have the form $a\boxtimes b$, where $a$ and $b$ are anyons of the two layers, respectively.

The second phase, QSL-II, is a single $\mathbb Z_2$ topological order, which can be obtained from the QSL-I phase by condensing the anyon $m\boxtimes m$, as shown in Fig.~\ref{fig:fig1} (d).
This condensation has two consequences:
First, it means that the visons on the two layers, $m\boxtimes1$ and $1\boxtimes m$, are identified as the same type of anyons, and become the vison ($m$ anyon) in the $\mathbb Z_2$ topological order. Second, the condensation of $m\boxtimes m$ confines the spinons on each layer, which are denoted  $e\boxtimes1$ and $1\boxtimes e$, because they have nontrivial mutual braiding statistics with $m\boxtimes m$.
On the other hand, the bound state $e\boxtimes e$ is still deconfined after the condensation, and becomes the $e$ anyon in the $\mathbb Z_2$ topological order.
We note that unlike the spinons in the QSL-I phase, the $e$ anyon in the QSL-II phase does not carry a fractional spin because it is a bound state of spinons on each layer.
As a result, a further condensation of $e$ anyons brings the QSL-II phase to a trivial paramagnetic phase without topological order or spontaneous symmetry breaking, which is the singlet phase in the phase diagram.

\begin{figure}[htp!]
	\centering
	\includegraphics[width=\columnwidth]{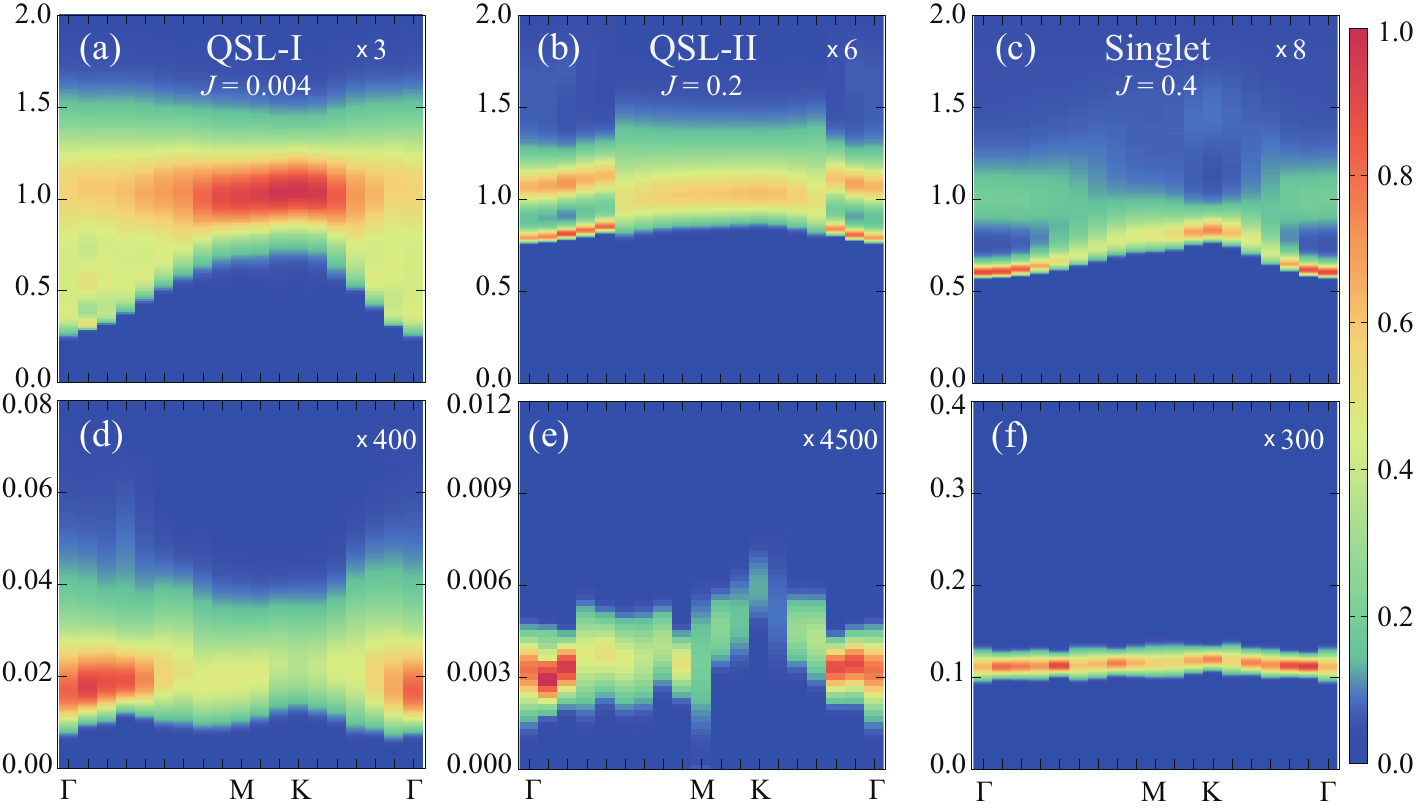}
	\caption{$S^{\pm}(\mathbf{q},\omega)$ spectra along the high symmetry path with $J_{\pm}=0.06$. (a) $J=0.004$ inside the QSL-I, (b) $J=0.2$ inside the QSL-II and (c)  $J=0.4$ inside the Singlet phases, with $\beta=600$ and the system size is $L=18$. $S^{zz}(\mathbf{q},\omega)$ spectra along the high symmetry path with the same parameter sets (d) inside the QSL-I phase, (e) inside the QSL-II phase and (f) inside the Singlet phase.}
	\label{fig:fig4}
\end{figure}

To reveal this theoretical understanding of the vestigial anyon condensation process, we designed the measurement of Wilson loops $W_A$ as shown in Fig.~\ref{fig:fig3} (a). $W_A=\langle \prod_{i \in M} 2 S_{i}^z \rangle$ measures the probability of moving an $m\boxtimes1$ anyon along loop $M$ on one of the kagome planes. The results are shown in Fig.~\ref{fig:fig3} (c); in a semi-log plot versus $L_{loop}$ (proportional to the perimeter), the perimeter-law decay of $W_A$ is present in both QSL-I and QSL-II, indicating the deconfinement of such anyons, and in the singlet and FM phases, the $W_A$ decays via an area-law, indicating the confinement of the anyons~\cite{Gregor2011} .

To distinguish QSL-I and QSL-II, we further designed another Wilson loop $W_B$ shown in Fig.~\ref{fig:fig3} (b). It is a vison loop that is half-way in the upper layer and half-way in the lower layer. As illustrated in Fig.~\ref{fig:fig3}(d), $W_B$  only exhibits perimeter-law decay in the QSL-II phase, indicating that the visons in the two layers belong to the same type. In other words, the vison in one layer is able to hop to another layer. In the QSL-I phase, on the other hand, $W_B$ decays via an area law, indicating that visons in the two layers belong to different types and they cannot hop from one layer to another. Therefore, $W_B$ does not form a closed Wilson loop in this phase. The area-law decays are also present in the singlet and FM phases.

With the establishment of the vestigial anyon condensation, we now discuss the phase transitions. As mentioned above, the continuous transition between QSL-I and QSL-II is driven by the condensation of $m\boxtimes m$.
This transition reduces the topological order from $\mathbb Z_2\boxtimes\mathbb Z_2$ to $\mathbb Z_2$ but does not break any global symmetry because $m\boxtimes m$ carries no symmetry fractionalization.
Therefore, it belongs to the $(2+1)$D Ising${}^*$ universality class, but since there is no local order parameter one cannot perform finite-size scaling to extract the correlation length exponent $\nu$. As for the specific exponent $\alpha$, in principle it can be deduced from the second derivative of the energy curves but that would require a parameter grid much finer than our current computational capability allows. Instead, the nature of this continuous transition is revealed from the continuous curve of the kinetic energy density as shown in Fig.~\ref{fig:fig2} and Fig.~\ref{fig:figs2}  and the analysis of the anyon condensation process in Fig.~\ref{fig:fig1} (d) (in that language, this can be deduced from an exact duality mapping to the transverse-field Ising model on the square lattice~\cite{Kogut1979}). Similarly, the transition from QSL-II to the singlet phase is driven by the condensation of the $e$ anyon, and also belongs to the Ising${}^*$ class.
The continuous transition from the QSL-I phase to the FM phase is driven by the condensation of spinons ($e\boxtimes1$ and $1\boxtimes e$), which eliminates the topological order completely and breaks the U(1) global symmetry, this is denoted the 3D $XY^*$ transition and we present data to illuustrate this understanding (this has been revealed in previous studies~\cite{Isakov2012,YCWang2018,YCWang2020}) in Fig.~\ref{fig:figs5}, in which we performed finite-size scaling of the spin stiffness across the transition, and the data collapse nicely reveals the correlation length exponent of $\nu=0.67$ for this transition. The transition from the QSL-II phase to the FM phase, however, is first order, consistent with the fact that there is no anyon in the QSL-II phase that carries a fractional U(1) charge.
Finally, the continuous transition between the singlet and the FM phase is of the Landau type and belongs to the 3D $XY$ universality class~\cite{WanwanXu2019,WLJiang2019}.

\subsection{Spectra} Finally, we illustrate the dynamical signature of the vestigial anyon condensation via the spin spectra that can be probed from neutron scattering. As shown in Fig.~\ref{fig:fig4}, we compute the dynamical spin-spin correlation functions $S^{\pm}_{\alpha\beta}(\mathbf{q},\tau)=\langle S^{+}_{-\mathbf{q},\alpha}(\tau)S^{-}_{\mathbf{q},\beta}\rangle$ and $S^{zz}_{\alpha\beta}(\mathbf{q},\tau)=\langle S^{z}_{-\mathbf{q},\alpha}(\tau)S^{z}_{\mathbf{q},\beta}\rangle$ where $\mathbf{q}$ moves along the high-symmetry path in the Brillouin zone [Fig.~\ref{fig:fig1} (b)] and $\alpha$, $\beta$ stand for the site index of the six-site unit cell. From the stochastic analytic continuation process and by taking the trace of the site indices, we obtain the spectra $S^{\pm}(\mathbf{q},\omega)$ and $S^{zz}(\mathbf{q},\omega)$, with the former probing the spinon pair and the later probing the vison pair. The spectra in Figs.~\ref{fig:fig4} (a) and \ref{fig:fig4}(d) in the QSL-I phase are consistent with those in previous works~\cite{GYSun2018,Becker2018} and the spinon and vison continua are clearly visible, with the former acquiring a larger gap and wider spread in the frequency and the latter acquiring a much smaller gap and signature of translational symmetry fractionalization (a finite momentum minimum at point $M$). Going into the QSL-II phase in Figs.~\ref{fig:fig4} (b) and \ref{fig:fig4}(e), due to the confinement of the spinon, i.e., condensation of $m\boxtimes m$, the $S^{\pm}(\mathbf{q},\omega)$ loses its continuum and becomes a sharp triplon band with a big gap, above which there are multi-triplon bands. On the other hand, since the visons in the two layers are now identical, the $S^{zz}(\mathbf{q},\omega)$ can still detect their continua as shown in Fig.~\ref{fig:fig4} (e), with an even smaller gap and spread in energy than those in Fig.~\ref{fig:fig4} (d). With a further increase in $J$ to the singlet phase, both spectra are now sharp and present the typical $S=1$ triplon dispersion in an anisotropic singlet-product paramagnet. All of Fig.~\ref{fig:fig4} therefore demonstrates the dynamical signature of the vestigial anyon condensation.

\section{Discussion} In this work, we construct a concrete coupled kagome QSL model for anyon condensation, and solve it with unbiased QMC numerics. Our vestigial anyon condensation process from QSL-I to QSL-II and eventually to the trivial singlet phase, is fully consistent with the topological field theory analysis, and our dynamics spectra provide the experimental relevant signature for its detection. This work paves the way for investigation of anyon condensation in more realistic quantum many-body models and eventually real materials.

\section*{Acknowledgements} We thank Ya-Hui Zhang for inspirational conversation on the topic. Y.C.W. acknowledges  support from the NSFC under Grant No. 11804383, the NSF of Jiangsu Province under Grant No. BK20180637, and the Fundamental Research Funds for the Central Universities under Grant No. 2018QNA39. C.W. acknowledges support from the RGC of Hong Kong SAR China (Grant Nos. ECS 21301018 and GRF 11300819). Z.Y. and Z.Y.M. acknowledge support from the RGC of Hong Kong SAR China (Grant Nos. GRF 17303019 and 17301420), and MOST through the National Key Research and Development Program (Grant No. 2016YFA0300502).  Y.Q. acknowledges supports from MOST under Grant No. 2015CB921700, and from the NSFC under Grant No. 11874115.  We thank the Computational Initiative of the Faculty of Science at the University of Hong Kong and the Tianhe-1A, Tianhe-2, and Tianhe-3 prototype platforms at the National Supercomputer Centers in Tianjin and Guangzhou for their technical support and generous allocation of CPU time.
\bibliographystyle{apsrev4-1}
\bibliography{bilayer-bfg}
\end{document}